\documentclass[10pt,a4paper,twoside]{article}
\usepackage{epsfig}
\usepackage{baltlat6}
\usepackage{array}
\usepackage{here}
\begin{document}
\ \
\vspace{0.5mm}
\setcounter{page}{1}
\vspace{8mm}


\titleb{GAS MOTIONS MAPPING FOR THREE SEYFERT GALAXIES}

\begin{authorl}
\authorb{A. Smirnova,}{1}
\authorb{A. Moiseev,}{1}
\authorb{I. Katkov}{2}~and
\authorb{V. Afanasiev}{1}
\end{authorl}

\begin{addressl}
\addressb{1}{Special Astrophysical Observatory, Russian Academy of Sciences,\\  Nizhnii Arkhyz, Karachaevo-Cherkesskaya Republic, 369167 Russia;\\
ssmirnova@gmail.com}
\addressb{2}{Sternberg Astronomical Institute, Universitetskii pr. 13,\\ Moscow, 119992 Russia}
\end{addressl}


\begin{summary}
We report preliminary results of a kinematical study for three Seyfert galaxies selected from a sample of nearby active galactic nuclei observed using 3D spectroscopy. The observations were performed at the prime focus of the SAO RAS  6-m telescope   with  the integral-field  spectrograph  MPFS and with a scanning  Fabry-–P\'{e}rot interferometer,  installed on the multimode device SCORPIO. Based on these data, the  monochromatic  maps and velocity fields in  different emission lines  were constructed.  We have detected a nuclear outflow or ionized gas motions associated with a radio jet in all galaxies circumnuclear regions.
\end{summary}


\resthead{GAS MOTIONS MAPPING FOR THREE SEYFERT GALAXIES}
{A. Smirnova et al.}

\sectionb{1}{Observations and data analysis.}
We have studied the kinematics of the ionized gas and stellar component in three Seyfert galaxies using methods of 3D spectroscopy. The observations were performed at the prime focus of the SAO RAS 6-m telescope. The circumnuclear regions were observed with the integral-field spectrograph MPFS (Afanasiev et al. 2001). The large-scale kinematics of the ionized gas in the H$\alpha$ emission line was studied using the SCORPIO multimode focal reducer  (Afanasiev \& Moiseev 2005) operating in the mode of scanning Fabry-P\'{e}rot Interferometer (FPI). The integral-field spectrograph MPFS takes simultaneous spectra of 256 spatial elements arranged in the form of 16x16 square lenses array with a scale of 1 arcsec per spaxel. The wavelength interval included numerous emission lines of the ionized gas and absorption features of the stellar population. Based on these data, the monochromatic maps and velocity fields in different emission lines of the ionized gas were constructed. We used the ULySS software (Koleva et al. 2009) for fitting of the spectra of stellar population as well as for construction of the stellar velocity fields. The velocity fields were fitted by the model of a pure circular rotation using the modification of the `tilted-ring' algorithm employed earlier to study NGC 6104 (Smirnova et al.  2006) and Mrk 334 (Smirnova \& Moiseev  2010).

\sectionb{2}{Mrk 198}

Mrk 198 is a Sy2 spiral galaxy with a bright asymmetric spiral arm at North-East direction. Faint arc-like feature is seen at 60-70 arcsec northeast from the nucleus Mrk 198 in our deep image (Smirnova et al. 2010). This tidal envelope traces the event of probable past minor merging. The velocity fields in the Balmer emission lines seems to be in a good agreement with the model of a regular rotating thin disc (Figures 1, 2). The kinematics of stars shows the same behaviour. However, the motions of gas emitted in higher-excitation lines have significant deviations from the pure circular rotation. The map of residual velocities in the [OIII] emission line reveals the significant (from -70 to +130 km/s) non-circular motions along  PA$\approx$30$^{\circ}$. The most probable explanation of the peculiar kinematics is a jet or an outflow penetrating through the surrounding interstellar medium. Our large-scale FPI data for the H$\alpha$ emission line demonstrate residual velocities in the central part of the galaxy, and also in the spiral arm. Significant non-circular motions seen only within the 3 arcsec, at larger distances (including bright asymmetric spiral arm) residual velocities are close to zero.


\begin{figure}[!tH]
\vbox{
\centerline{\psfig{figure=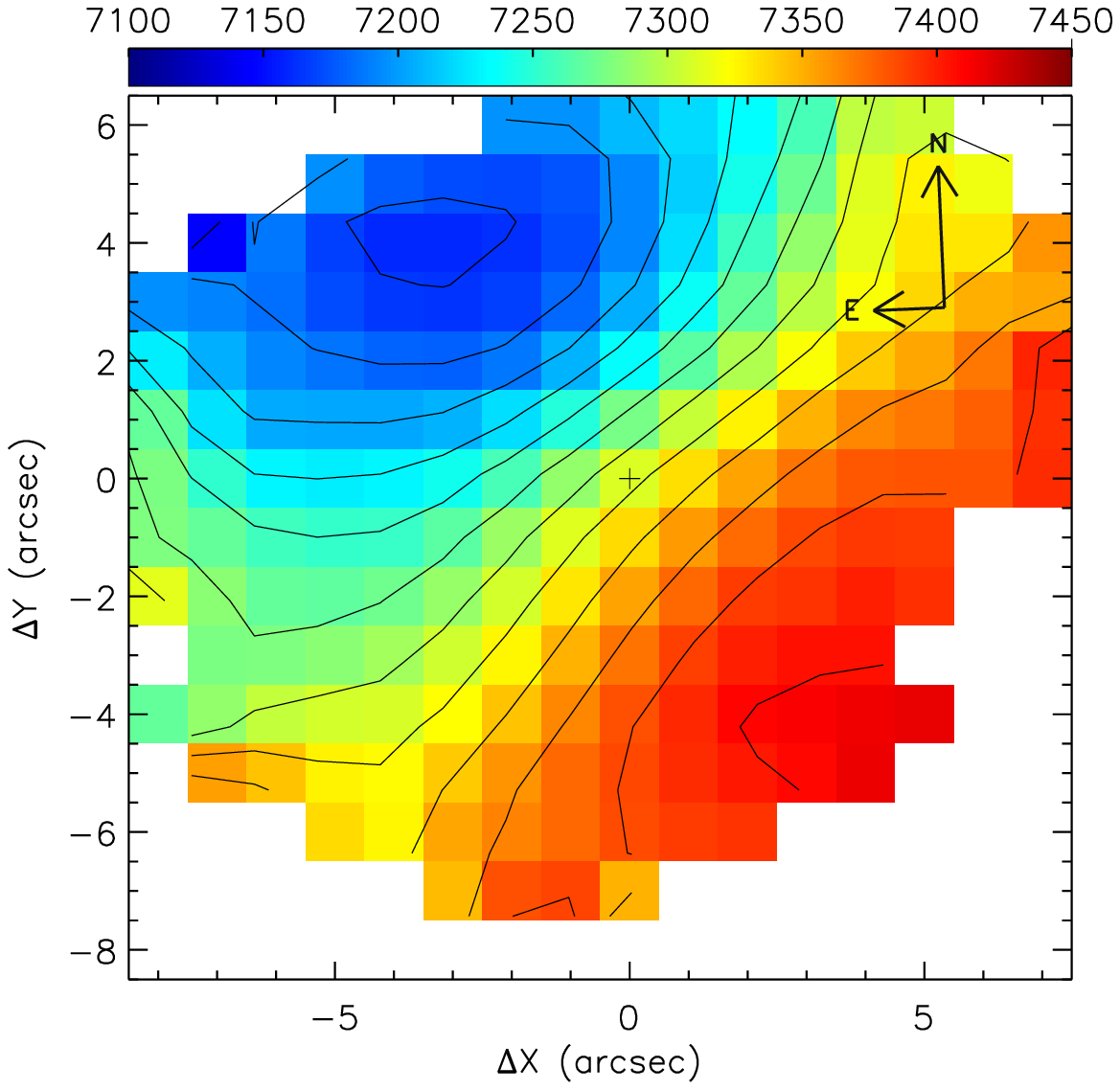,width=38mm,angle=0,clip=}\psfig{figure=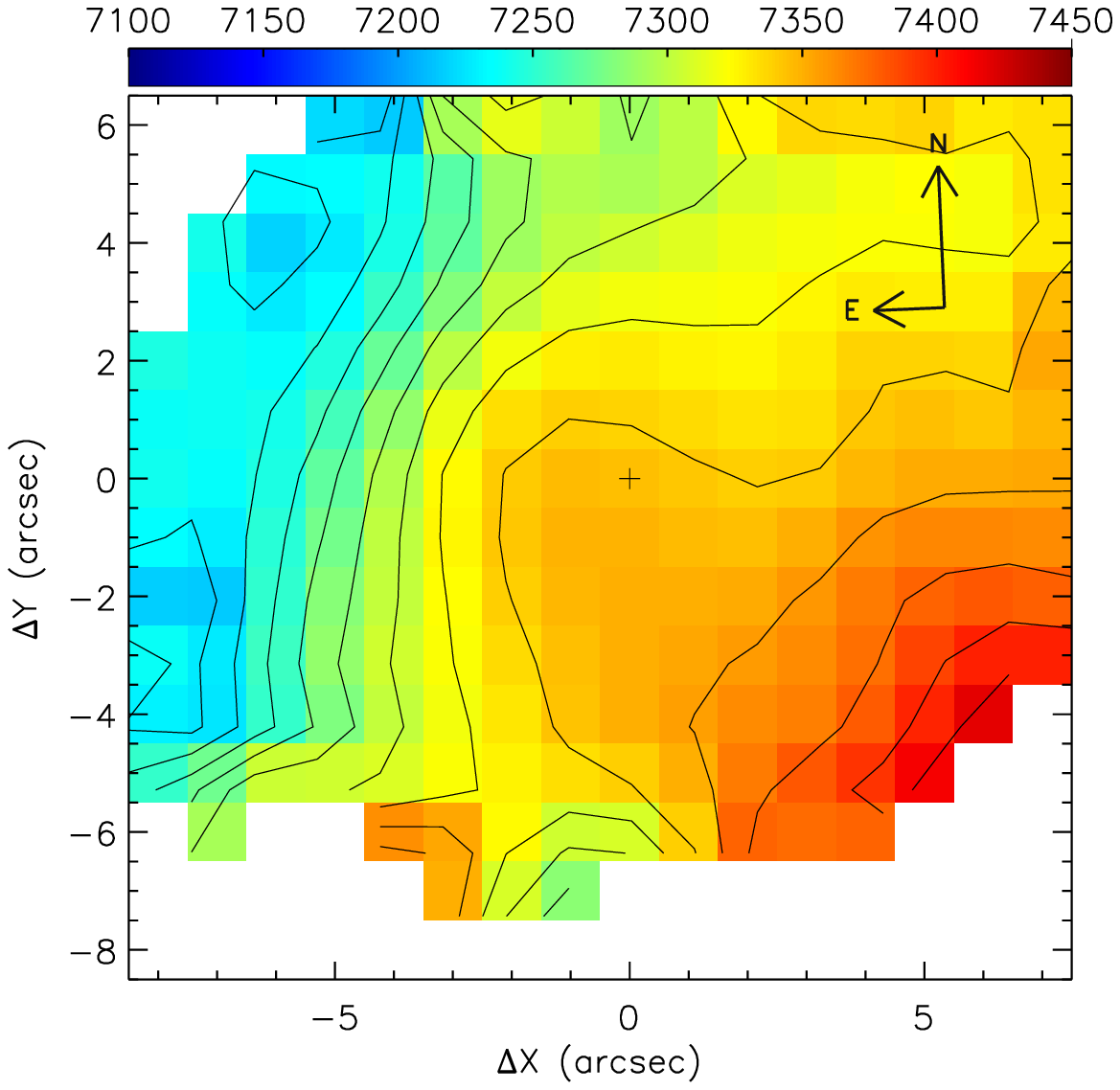,width=38mm,angle=0,clip=}\psfig{figure=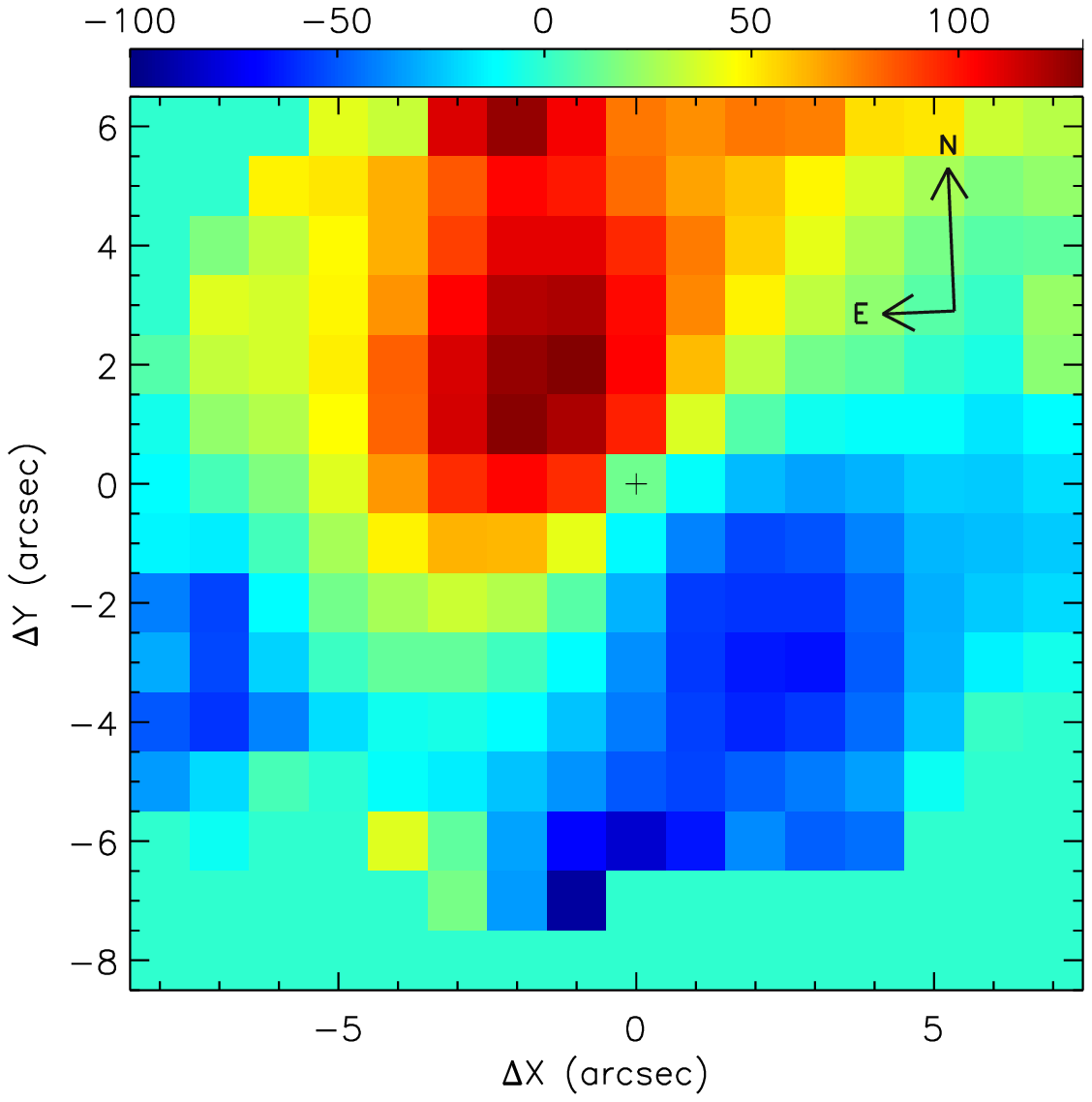,width=37mm,angle=0,clip=}}
\vspace{1mm}
\captionb{1}
{Mrk 198 MPFS data: velocity fields in the H$\alpha$ (left), [OIII] (middle) and [OIII] residual velocity map (right).}
}
\end{figure}

\begin{figure}[!tH]
\vbox{
\centerline{\psfig{figure=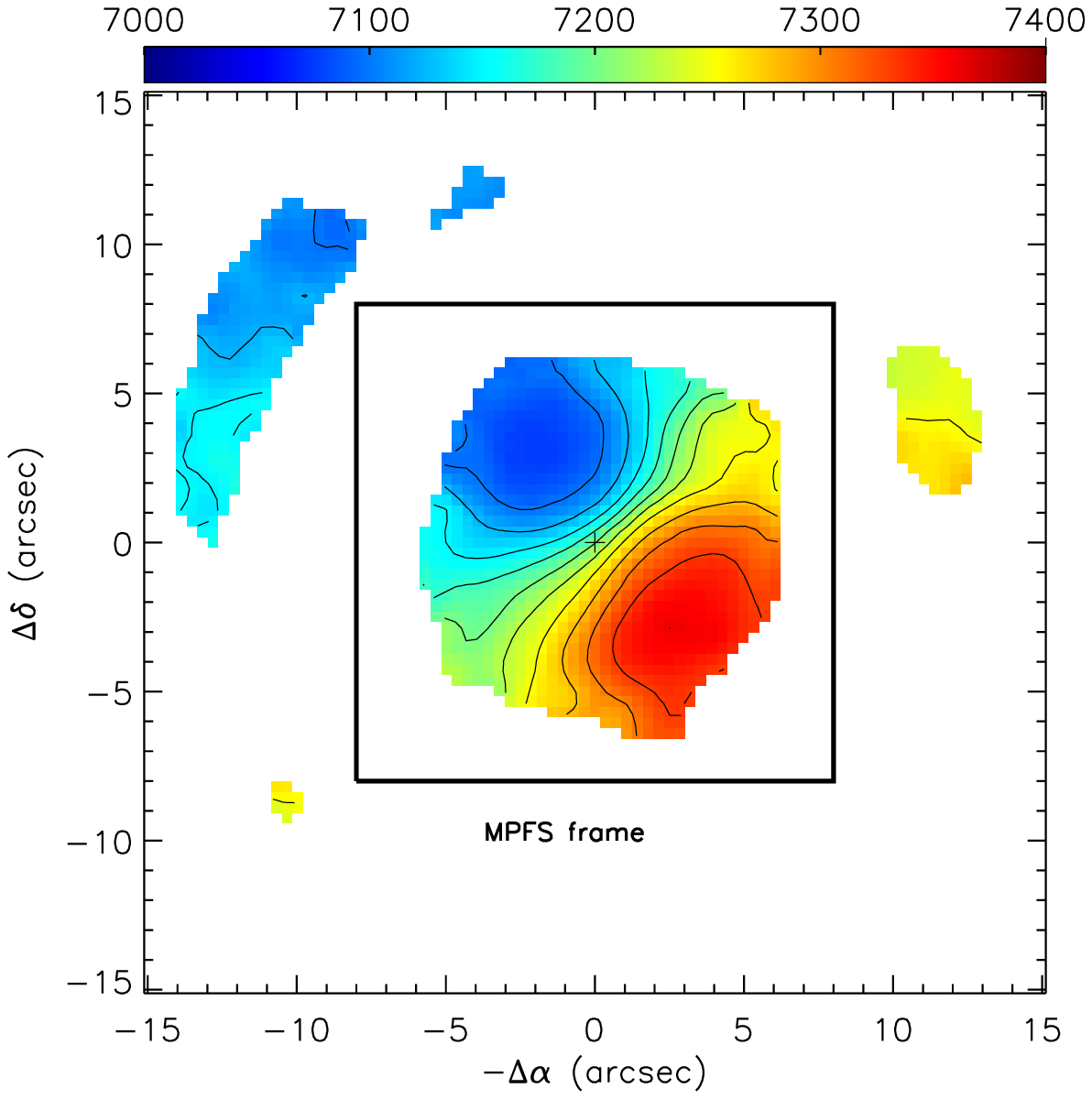,width=38mm,angle=0,clip=}\psfig{figure=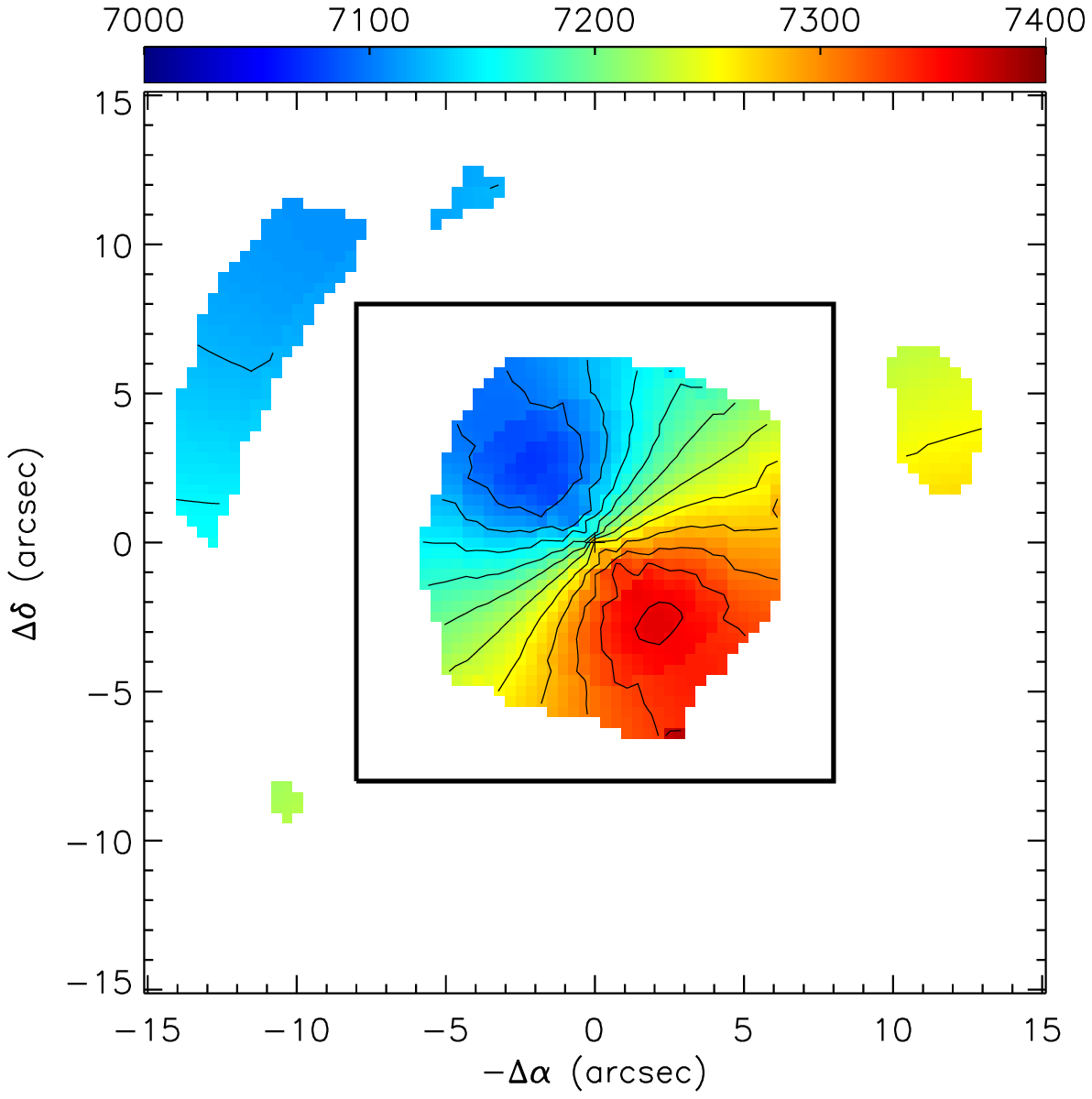,width=38mm,angle=0,clip=}\psfig{figure=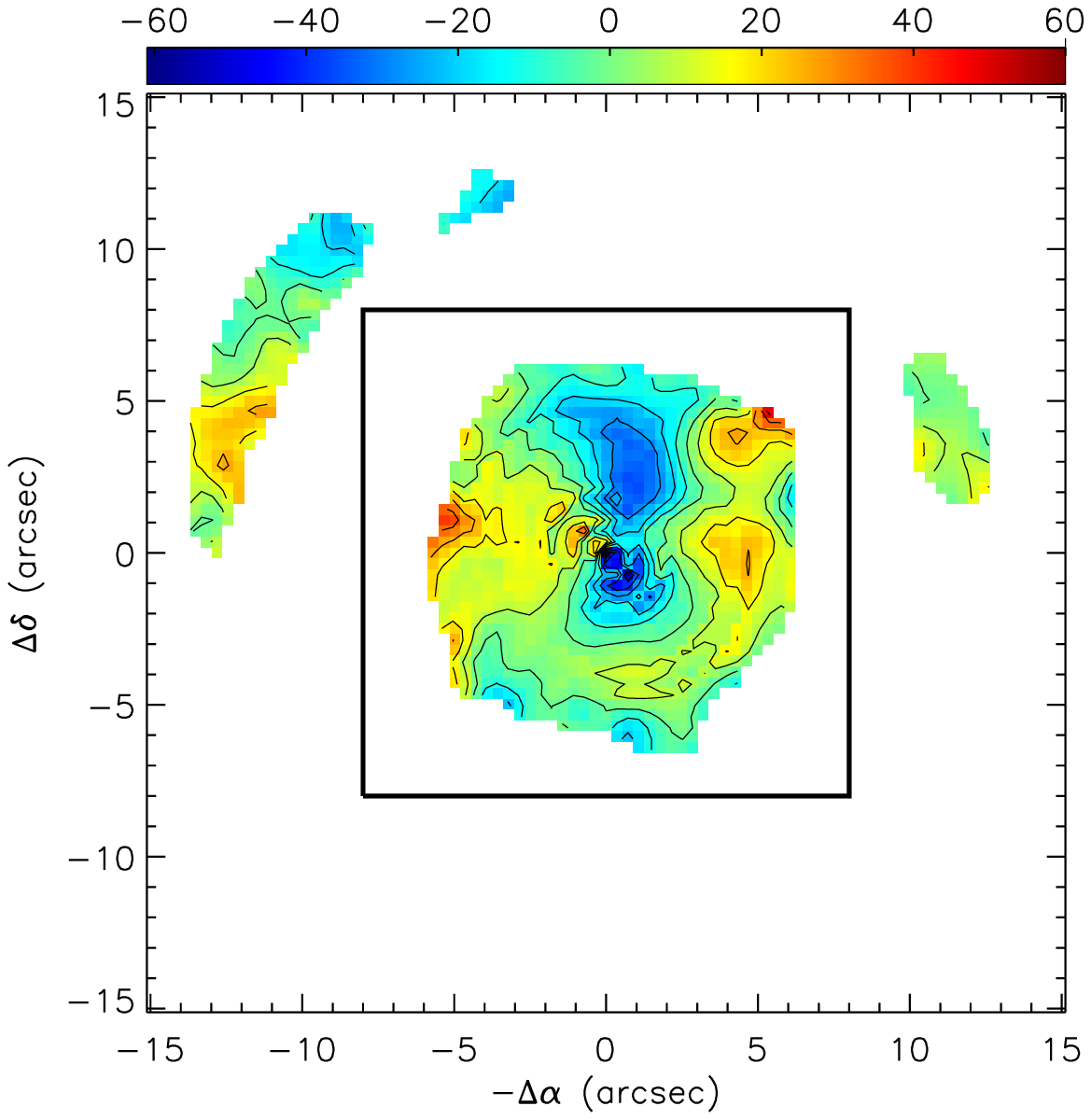,width=38mm,angle=0,clip=}}
\vspace{1mm}
\captionb{2}
{Mrk 198: The H$\alpha$ emission line velocity field derived from the FPI data (left), a titled-ring model of the circular rotation (middle) and the residual velocity map (right).}
}
\end{figure}

\sectionb{3}{Mrk 291}

Mrk 291 is a Sy2 galaxy, it has a bar and two spiral arms started at the end of the bar (Adams  1973). It is a strongly isolated object. The velocity field derived in the H$\alpha$ emission line demonstrates a regular rotation plus radial streaming motions along the stellar bar (two regions with non-zero residuals at r$>$5$''$, see Figure~3). In contrast to this picture, a significant excess of blueshifted velocities (about -100 km/s) in the [OIII] emission line is detected (Figure~4). Similar gas outflows associated with a nuclear starburst, or a jet-clouds interaction, or with a hot wind emerging from an active nucleus were already found in the integral field spectroscopy data for several Seyfert galaxies  (for references see Smirnova \& Moiseev 2010).

\begin{figure}[!tH]
\vbox{
\centerline{\psfig{figure=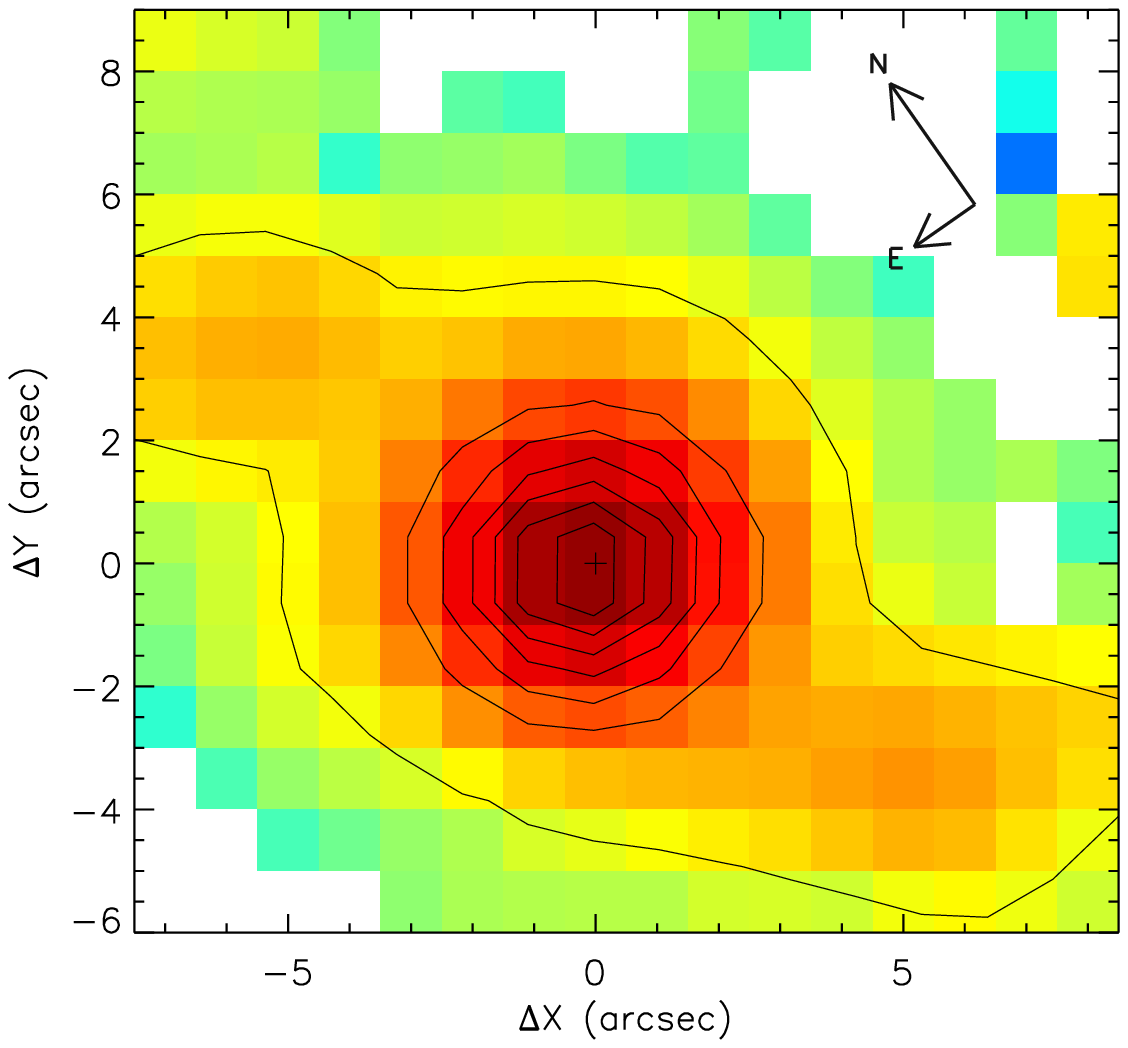,width=38mm,angle=0,clip=}\psfig{figure=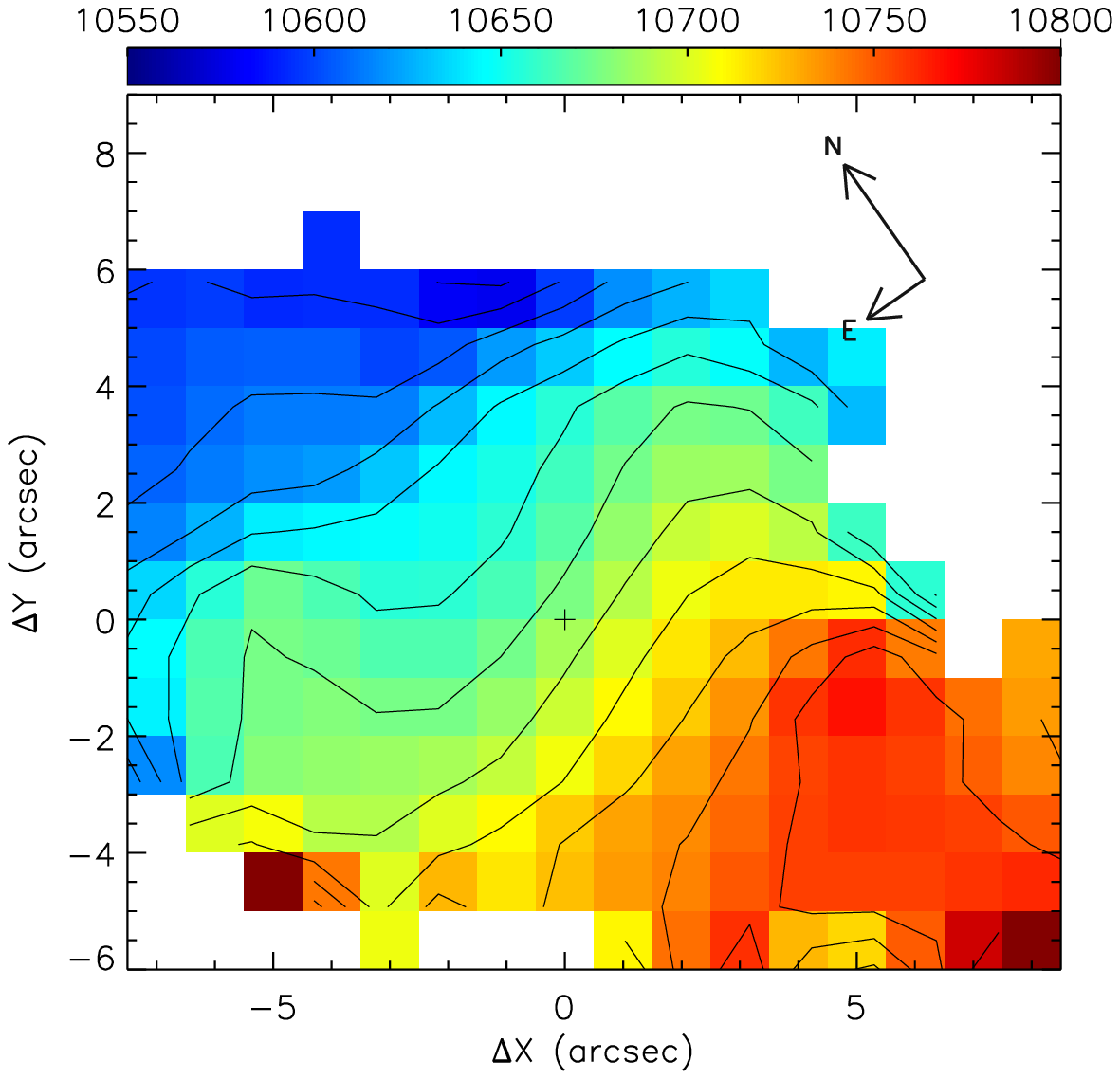,width=39mm,angle=0,clip=}\psfig{figure=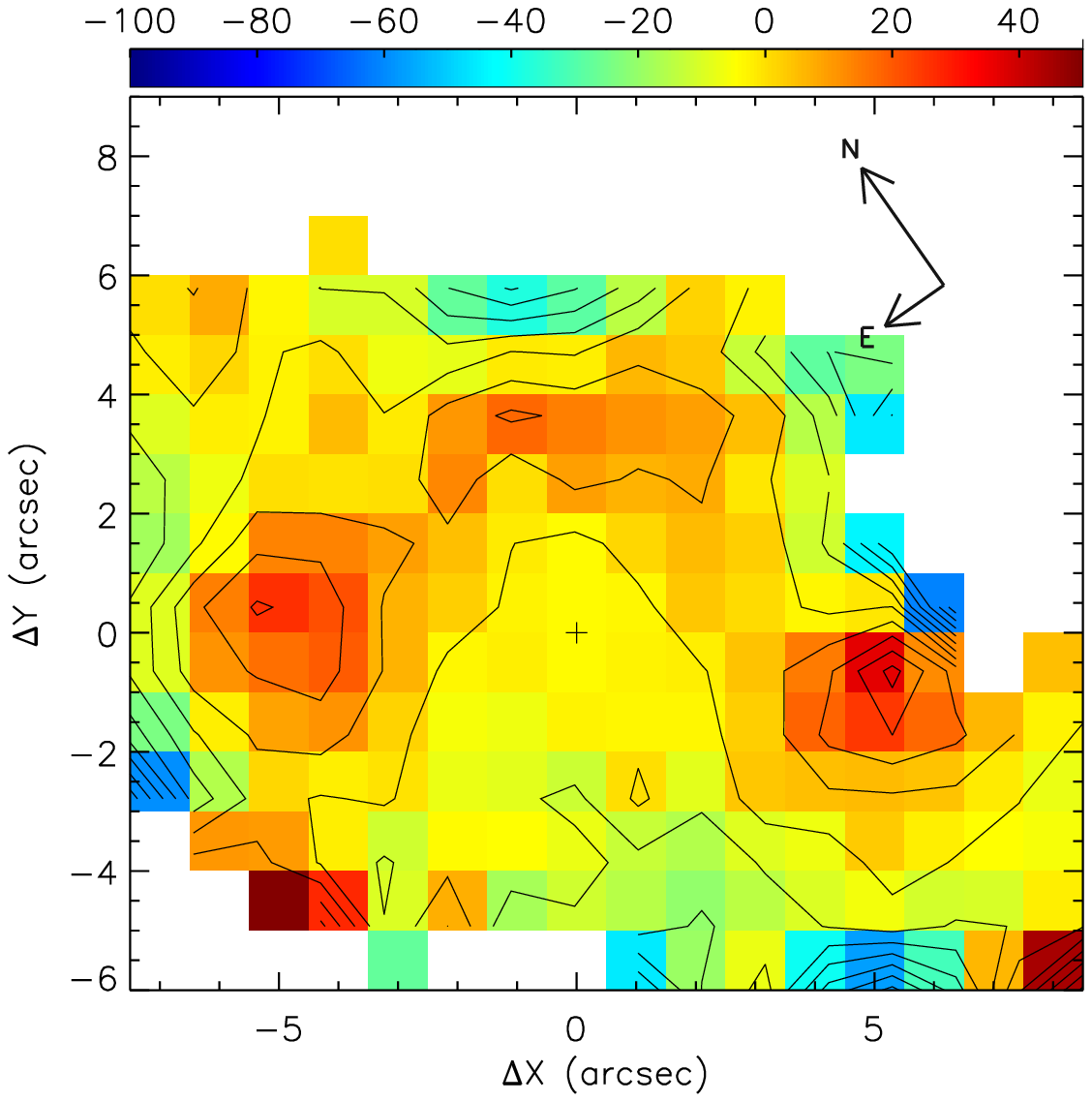,width=37mm,angle=0,clip=}}
\vspace{1mm}
\captionb{3}
{Mrk 291 MPFS data for the H$\alpha$: intensity map (left), velocity field (middle) and  the residual velocity map (right).}
}
\end{figure}

\begin{figure}[!tH]
\vbox{
\centerline{\psfig{figure=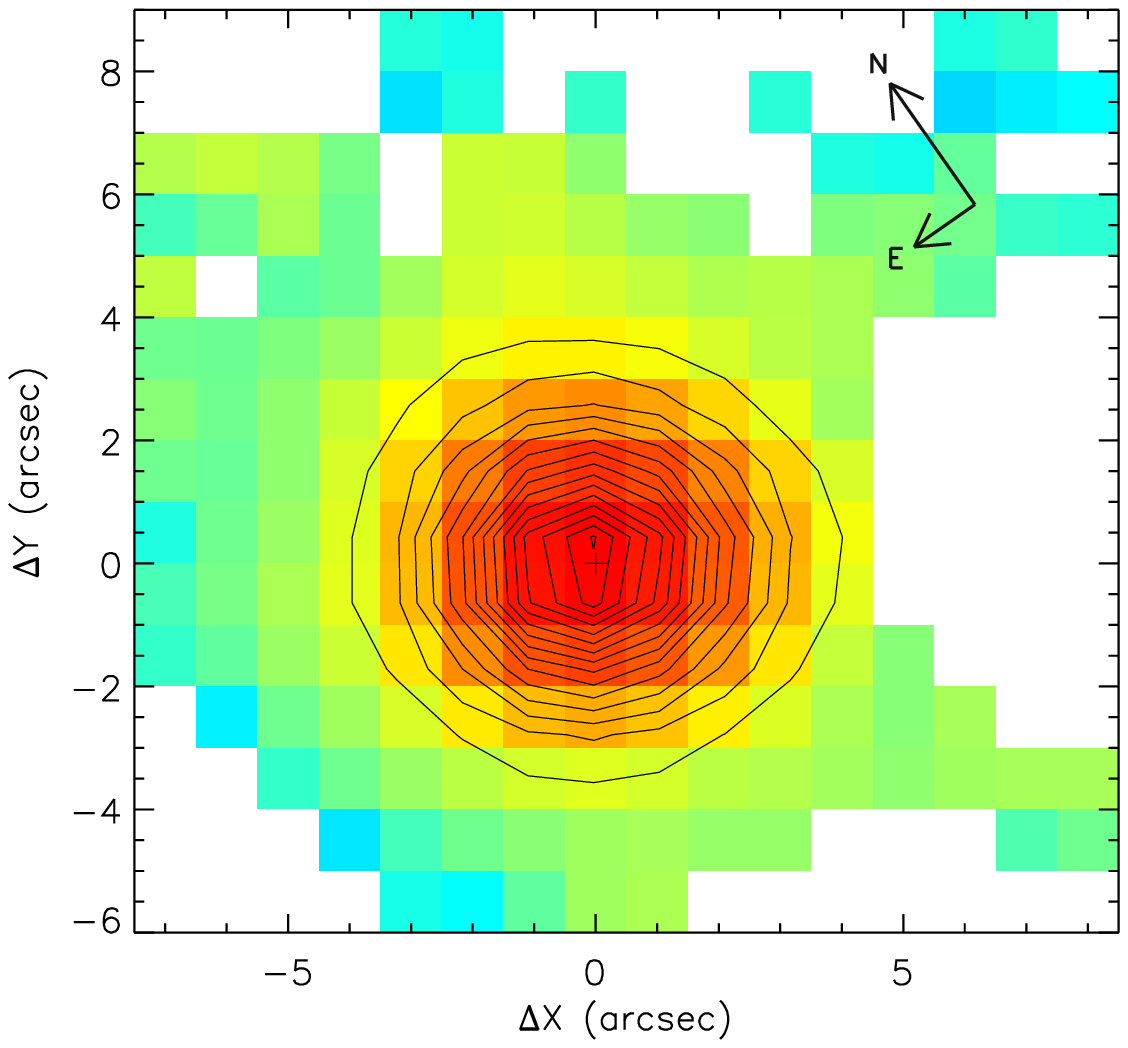,width=38mm,angle=0,clip=}\psfig{figure=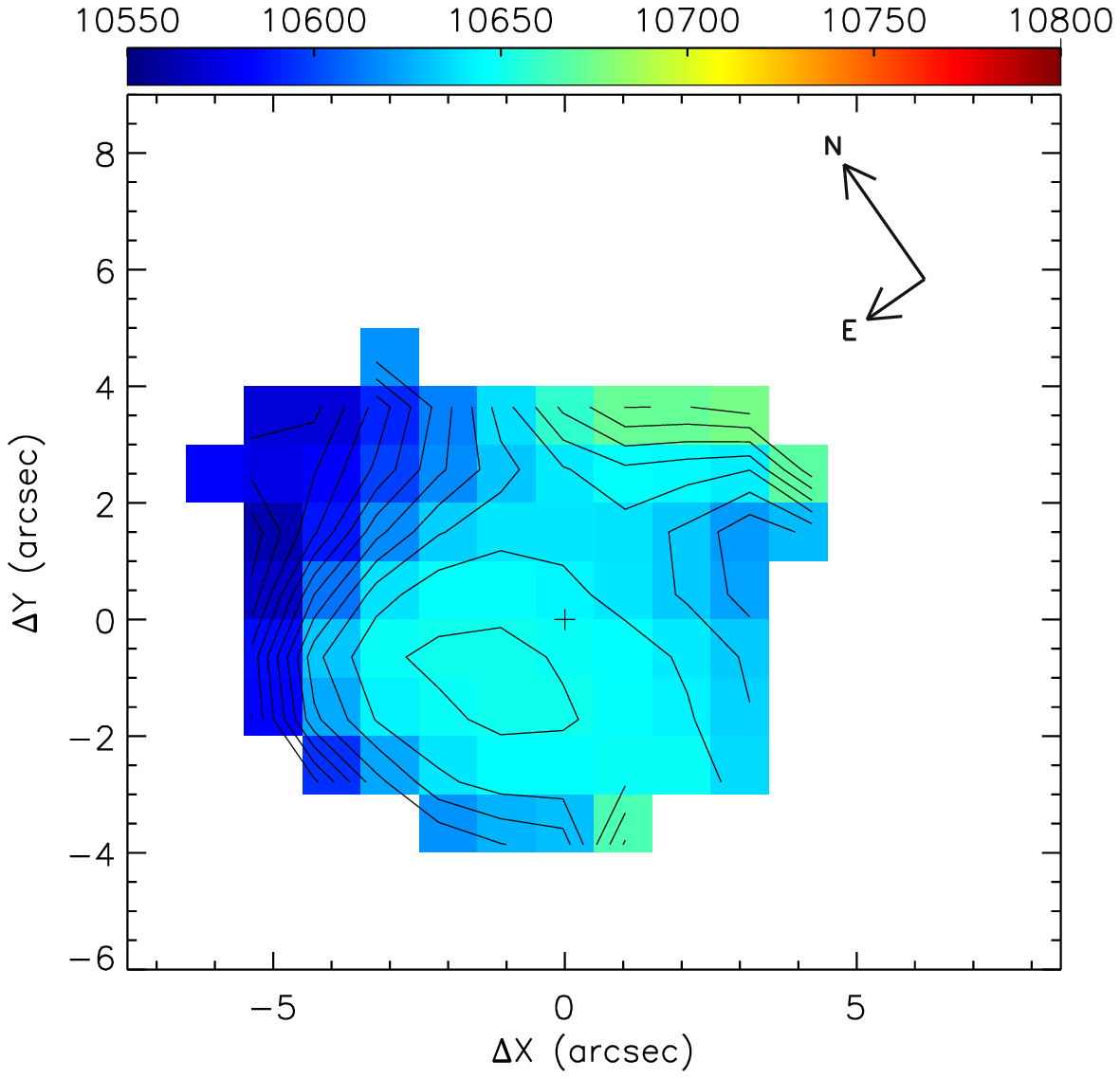,width=39mm,angle=0,clip=}\psfig{figure=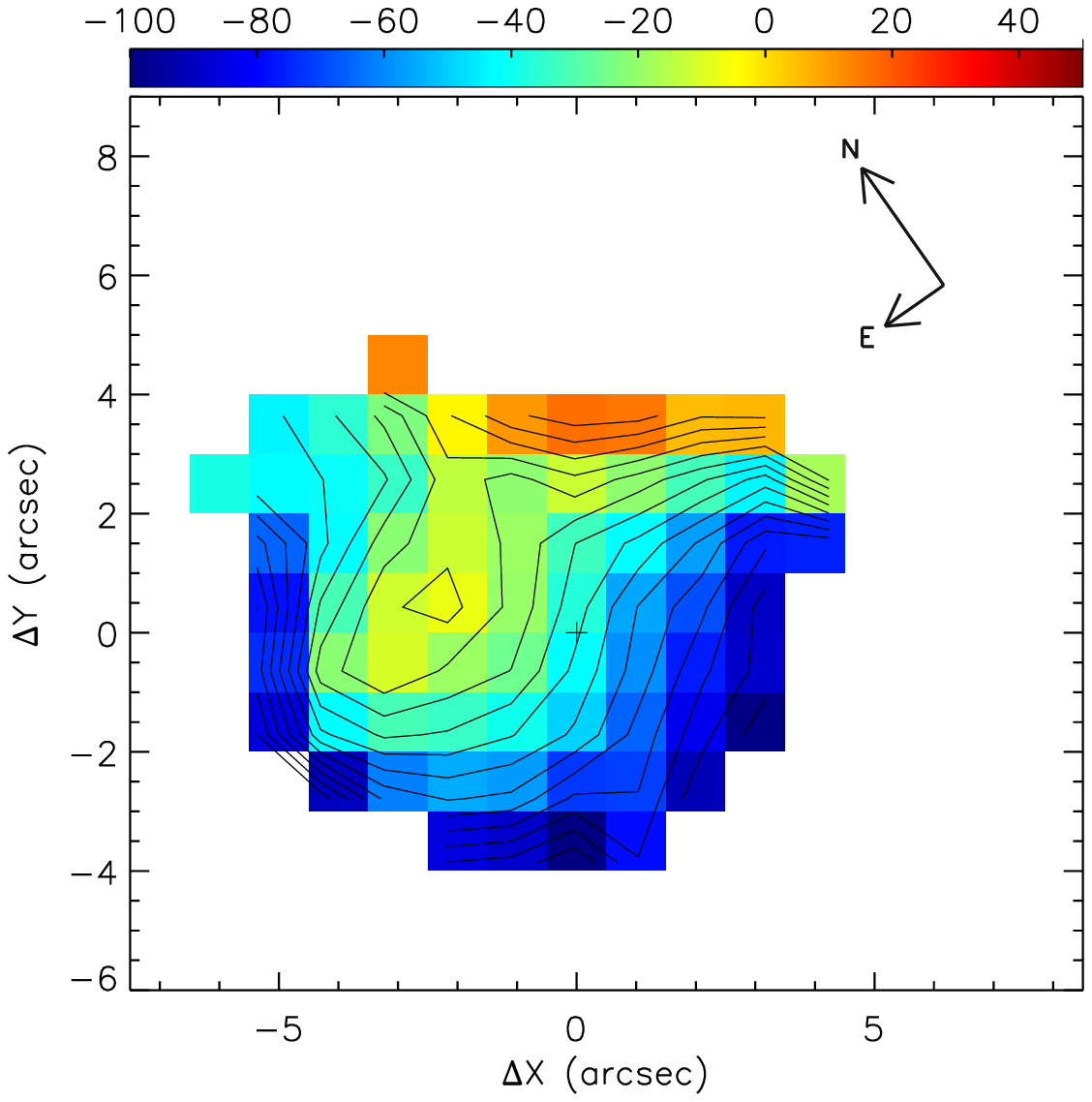,width=38mm,angle=0,clip=}}
\vspace{1mm}
\captionb{4}
{Mrk 291 MPFS data for the [OIII]: intensity map (left), velocity field (middle) and  the residual velocity map (right).}
}
\end{figure}

\sectionb{4}{Mrk 348}

Mrk 348 is a Sy2 spiral tidally-disrupted galaxy (Simkin et al. 1987). FPI data reveals extended spiral structure with a projected companion galaxy (Figure~5). Maps of the residual velocities in Balmer emission lines as well as in forbidden lines demonstrate significant non-circular motions ($\pm$50 km/s) at North and South direction from the nucleus (Figure~6). The radio image of the galaxy shows extended structure up to 0.2$''$  at P.A. =170$^{\circ}$ (Ulvestad \& Wilson  1984). According to the HST images (Capetti 2002; Falcke et al.  1998) the extended (up to r=1.5$''$)  [OIII] emission structure is elongated in the same direction. Thereby peculiar motions observed on the MPFS maps at the distances $r=1-4''$ are associated with a cocoon of the ionized gas surrounded the radio jet which is penetrate into interstellar medium. A similar combination (inner radio jet + velocities perturbations in the outer regions) was recently found in Mrk533 (Smirnova et al. 2007).

\begin{figure}[!tH]
\vbox{
\centerline{\psfig{figure=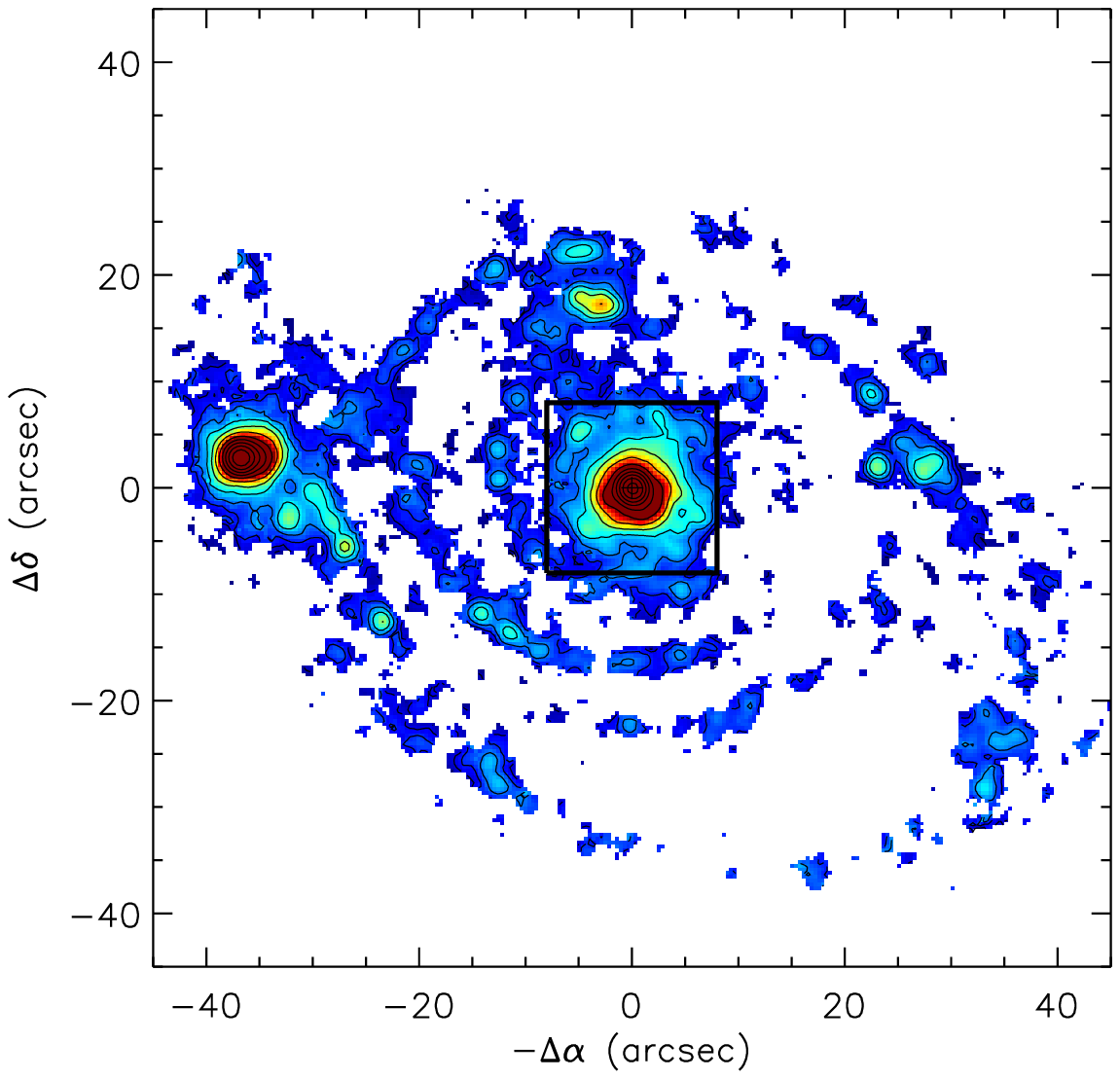,width=38mm,angle=0,clip=}\psfig{figure=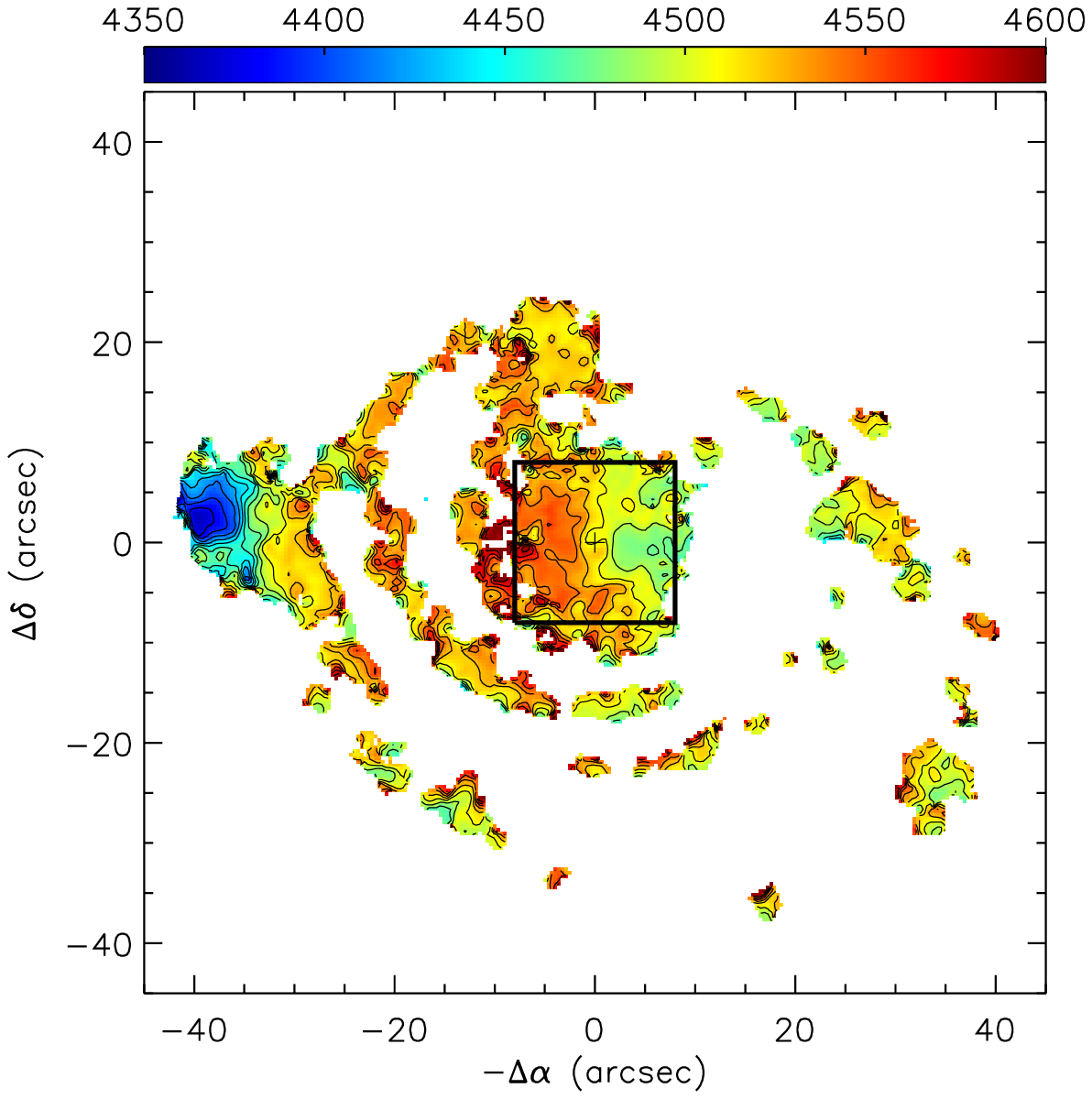,width=39mm,angle=0,clip=}\psfig{figure=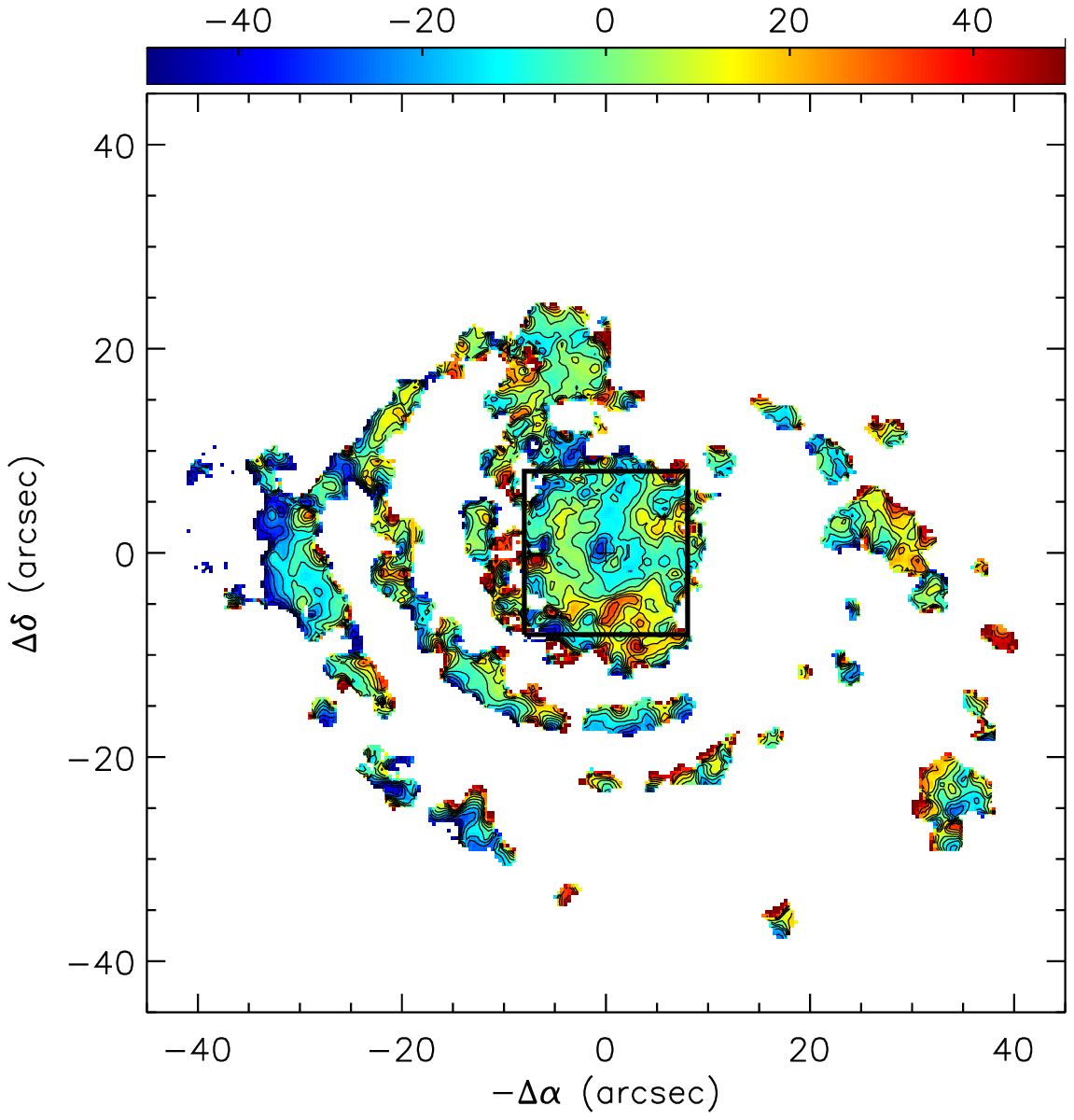,width=38mm,angle=0,clip=}}
\vspace{1mm}
\captionb{5}
{Mrk 348: the FPI data in H$\alpha$: monochromatic map (left), velocity field (middle) and the residual velocities (right).}
}
\end{figure}

\begin{figure}[!tH]
\vbox{
\centerline{\psfig{figure=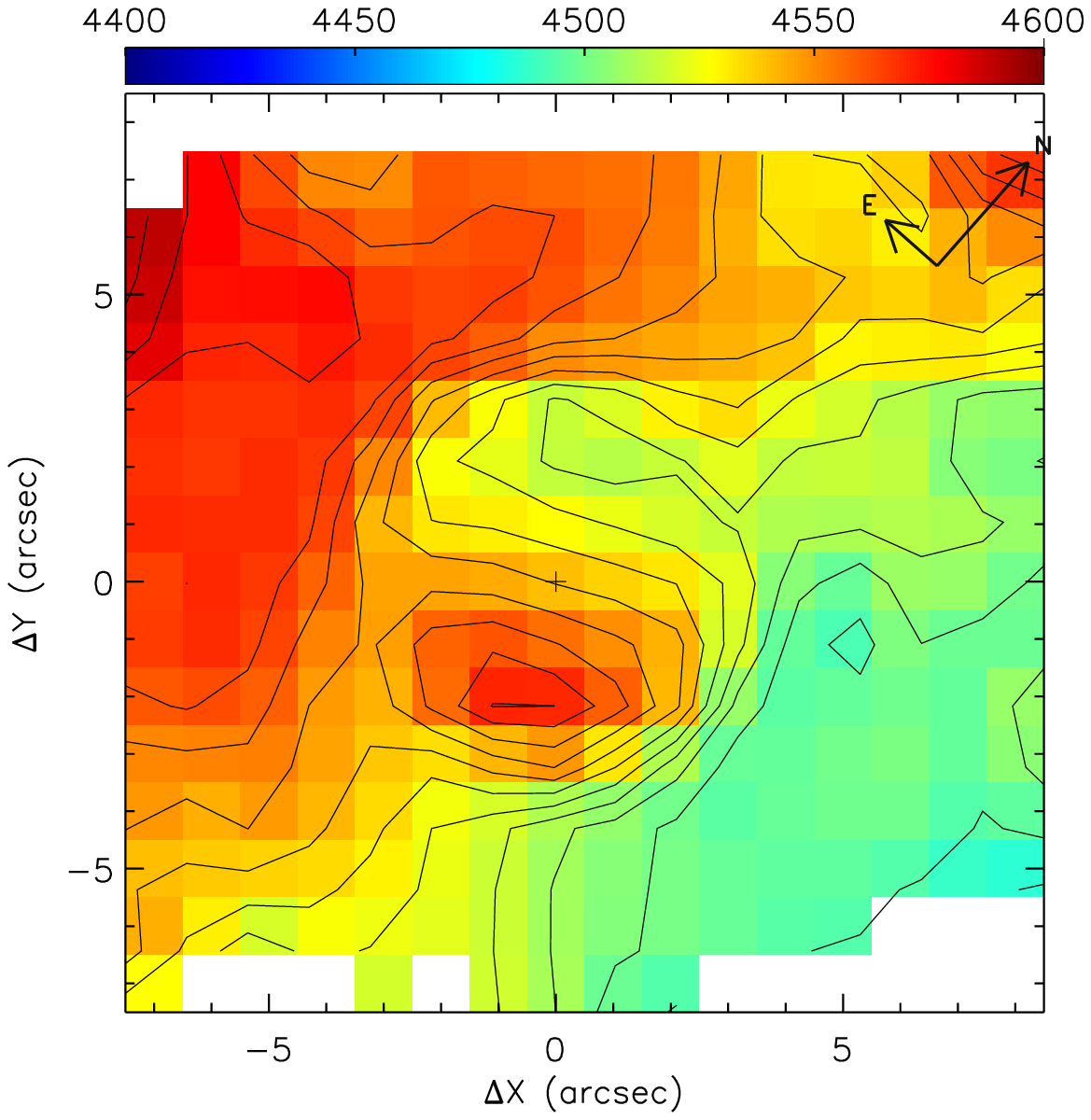,width=38.5mm,angle=0,clip=}\psfig{figure=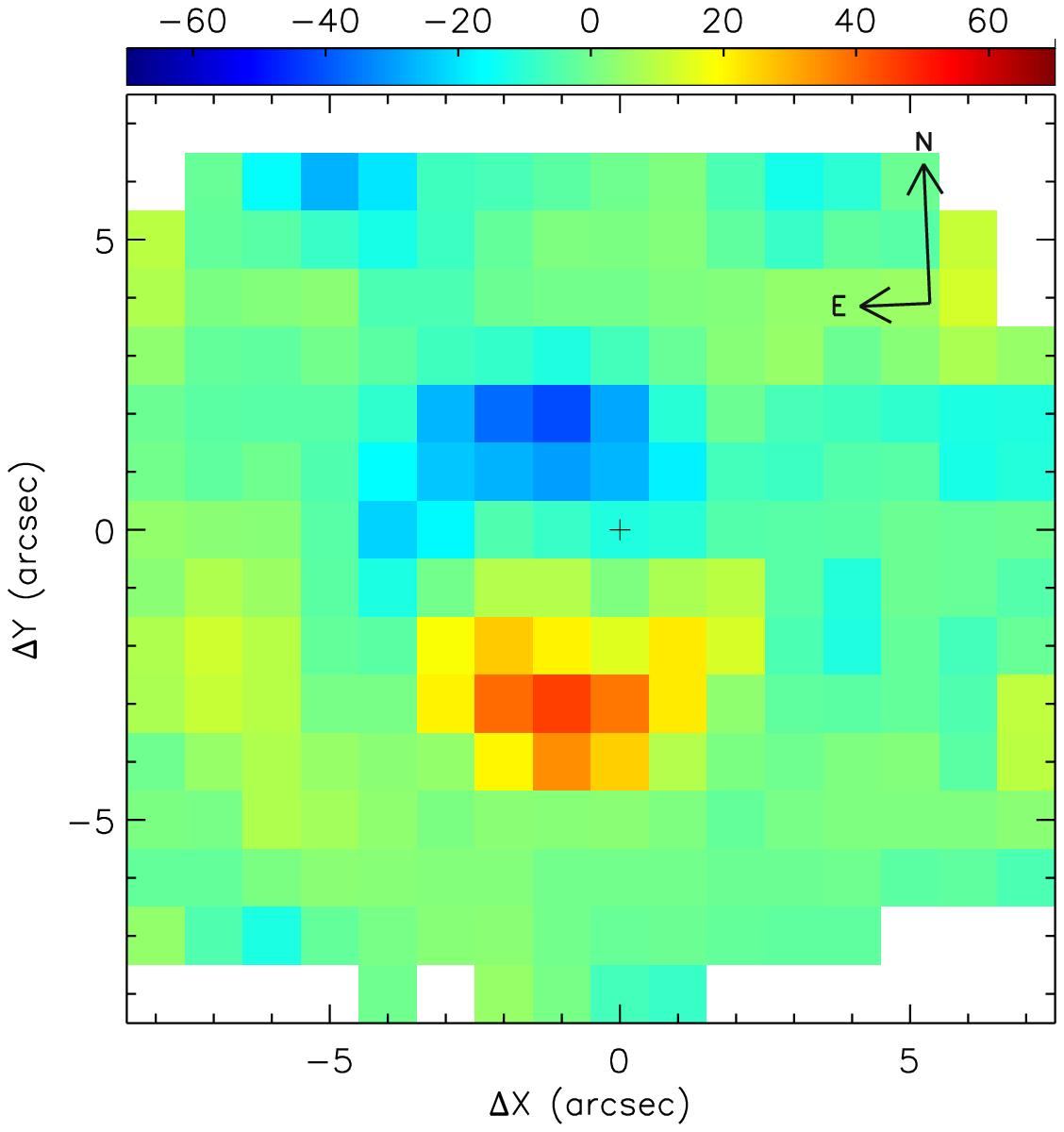,width=37mm,angle=0,clip=}\psfig{figure=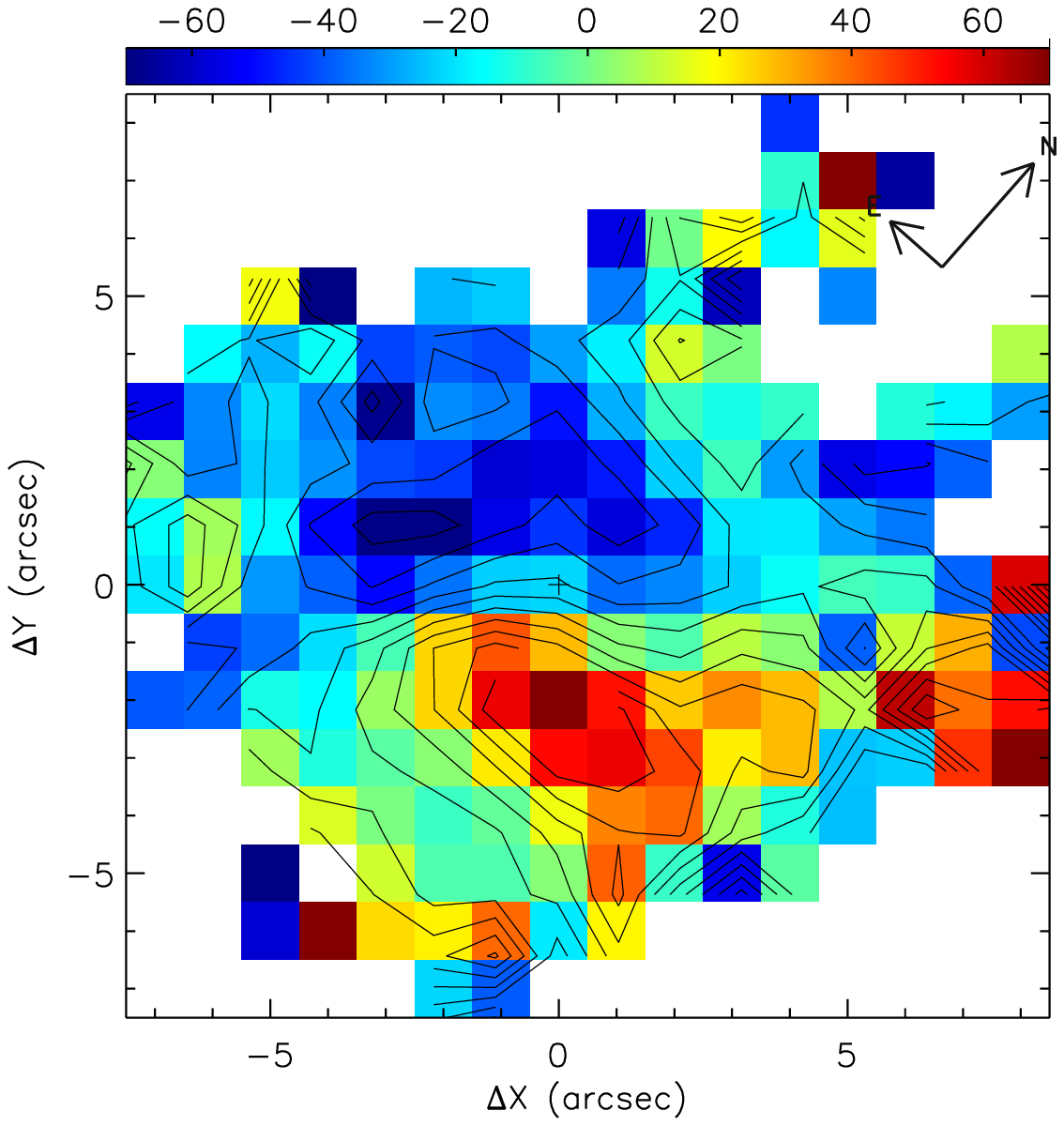,width=37.4mm,angle=0,clip=}}
\vspace{1mm}
\captionb{6}
{Mrk 348 MPFS data: velocity fields in the H$\alpha$ (left) and residual velocity map (middle), [OIII] residual velocity map (right).}
}
\end{figure}



\thanks{ We thank the Organizing Committee for financial support and efficient
organization of the conference. This work was supported by the Russian Foundation for Basic Research (project
no.~09-02-00870) and by the Russian Federal Program `Kadry' (contract no.~14.740.11.0800). A.M. is also grateful to the `Dynasty' Fund.}

\References

\refb Adams T. 1977, ApJS, v.33, p.19

\refb Afanasiev V.L.,  Dodonov S.N.,  Moiseev A.V. 2001, {\it in Stellar Dynamics: From Classic to Modern}, eds. Ossipkov
  L.~P. \&  Nikiforov I.~I., Saint Petersburg,  103

\refb Afanasiev V. L., Moiseev A. V.  2005, Astronomy Letters,  31,  193

\refb Capetti A.  2002, RevMexAA (Serie de Conferencias),  13, 163

\refb Falcke H.,  Wilson A.S., Simpson C. 1998, ApJ,  502,  199

\refb Koleva M., Prugniel Ph., Bouchard A., Wu Y. 2009,  A\&A,  501,  1269

\refb Simkin  S.,  van Gorkom  J., Hibbard  J., Su  H.-J.  1987, Science,  235,  1367

\refb Smirnova A.,  Gavrilovic N., Moiseev  A., Popovic  L., Afanasiev  V., Jovanovic  P.  2007, MNRAS,  377, 480

\refb Smirnova A. A.,  Moiseev A. V.  2010, MNRAS,  401,  30

\refb Smirnova A. A., Moiseev  A. V., Afanasiev  V. L. 2006, Astronomy Letters,  32,  520

\refb Smirnova A. A.,  Moiseev A. V., Afanasiev  V. L.  2010, MNRAS,  408,  400

\refb Ulvestad J., Wilson A. 1984, ApJ,  285,  439

\end{document}